\documentclass[pre,graphicx,showpacs,superscriptaddress]{appolb}

\usepackage{graphicx}
\usepackage[francais,english]{babel}
\usepackage[latin1]{inputenc}
\usepackage[T1]{fontenc}
\usepackage{amsmath}
\usepackage{amssymb}
\usepackage{color}
\usepackage{psfrag}
\usepackage{graphics}
\usepackage{mathrsfs}
\usepackage{verbatim}
\usepackage{textcomp}
\usepackage{xspace}

\newcommand{\nn}{\nonumber}

\newcommand{\rmd}{{\mathrm{d}}}

\newcommand{\lmax}{\lambda_{\rm max}}

\begin{document}
\title{The density of eigenvalues seen from the soft edge of random matrices 
in the Gaussian $\beta$-ensembles \thanks{for Random Matrix Theory : Foundations and Applications, Krak\`ow.}}

\author{Anthony Perret, Gr\'egory Schehr
\address{Univ. Paris-Sud -- Paris 11, CNRS, Laboratoire de Physique Th\'eorique et Mod\`eles Statistiques (LPTMS), 91405 Orsay Cedex, France}}

\date{\today}

\maketitle
\begin{abstract}
We characterize the phenomenon of ``crowding'' near the largest eigenvalue $\lmax$ of random $N \times N$ matrices belonging to the Gaussian $\beta$-ensemble of random matrix theory, including in particular the Gaussian orthogonal ($\beta=1$), unitary ($\beta=2$) and symplectic ($\beta = 4$) ensembles. We focus on two distinct quantities: (i) the density of states (DOS) near $\lmax$, $\rho_{\rm DOS}(r,N)$, which is the average density of eigenvalues located at a distance $r$ from $\lmax$ (or the density of eigenvalues seen from $\lmax$) and (ii) the probability density function of the gap between the first two largest  eigenvalues, $p_{\rm GAP}(r,N)$. Using heuristic arguments as well as well numerical simulations, we generalize our recent exact analytical study of the Hermitian case (corresponding to $\beta = 2$). We also discuss some applications of these two quantities to statistical physics models.    
\end{abstract}

\section{Introduction}


During the last 20 years, there has been an important activity, both in mathematics and in physics, aiming at describing the fluctuations of the largest eigenvalue  in ensembles of random matrices \cite{TW94a,TW96,ben,satyadean1,satyadean2,majverg,MS14}. The most studied ones in this context, which we also focus on here, are probably the Gaussian $\beta$-ensembles where the joint probability density function (PDF) of the $N$ real eigenvalues $\lambda_1, \cdots, \lambda_N$ is given by: 
\begin{eqnarray}\label{jPDF}
P_{\rm joint}(\lambda_1,\lambda_2,...,\lambda_N)=\frac{1}{Z_N}\prod_{i<j}|\lambda_i-\lambda_j|^\beta \, \exp\hspace{-0.05cm}\left({-\frac{\beta}{2}\sum_{i=1}^{N}\lambda_i^2}\right) \;,
\end{eqnarray}
where the normalization constant is $Z_N = (2\pi)^{N/2} \beta^{-N/2-\beta N(N-1)/4} \Gamma(1+\beta/2)^N\prod_{j=1}^N \Gamma(1+\beta j/2)$ and where $\beta > 0$ is the Dyson index that can take any real value. The classical values correspond to $\beta = 1, 2$ and $4$, associated respectively to 
the Gaussian Orthogonal Ensemble (GOE), the Gaussian Unitary Ensemble (GUE) and the Gaussian Symplectic Ensemble (GSE). Note that for arbitrary $\beta$, it is possible to associate a matrix model to (\ref{jPDF}) (namely tridiagonal random matrices introduced in~\cite{DE02}). The fluctuations of the largest eigenvalues $\lmax=\max_{1 \leq i \leq N} \lambda_i$, characterized by its cumulative distribution $F_N(y) = {\rm Proba.}[\lambda_{\max} \leq y]$, are now well understood. Indeed, we have now a precise characterization of the typical fluctuations of $\lmax$, when $|\lmax - \sqrt{2N}|$ is of order ${\cal O}(N^{-1/6})$, which are described the TW distributions \cite{TW94a,TW96} as well as of the large deviations of $\lambda_{\max}$, when $|\lmax - \sqrt{2N}|$ is of order ${\cal O}(\sqrt{N})$, where $F_N(y)$ is described by large deviations functions \cite{ben,satyadean1,satyadean2,majverg,MS14} (both left and right tails).



\begin{figure}[h!]
\begin{center}
\resizebox{9cm}{!}{\includegraphics{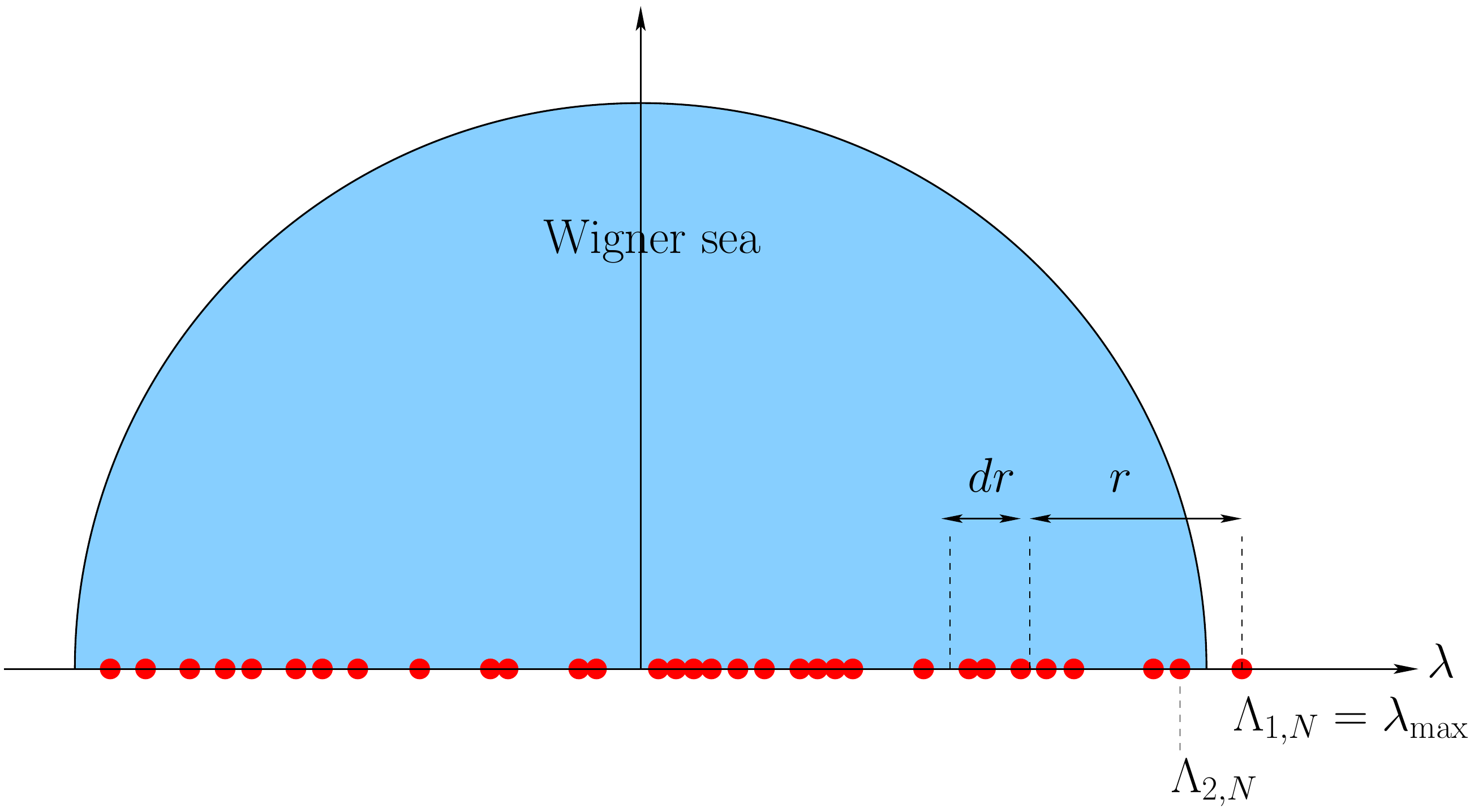}}
\caption{Different quantities characterizing the "crowding" near the largest eigenvalue $\lmax$ studied in this paper: (i) the mean density of states $\rho_{\rm DOS}(r,N)$ such that $\rho_{\rm DOS}(r,N) \rmd r$ is the mean number of eigenvalues located in the interval $[\lmax-r-\rmd r,\lmax-r]$ and (ii) the PDF $p_{\rm GAP}(r,N)$ of the spacing between the two largest eigenvalues, $p_{\rm GAP}(r,N) \rmd r = {\Pr.} [(\Lambda_{1,N} - \Lambda_{2,N}) \in [r, r+\rmd r]]$.}
\label{fig_illust}
\end{center}
\end{figure}

However in several situations, some observables related to such spectral statistics (\ref{jPDF}), might be sensitive not only to $\lmax$ but also to 
near-extreme eigenvalues, whose amplitudes are smaller but close to this largest eigenvalue. This general feature, not restricted to random matrices, 
has led physicists to study the phenomenon of ``crowding'' close to the maximum \cite{SM2007}. This was studied in detail for the case of  independent
and identical random variables \cite{SM2007} and more recently for Brownian motion \cite{PCMS13}. In Ref. \cite{PS14} we proposed to characterize this phenomenon of ``crowding'' in random matrix models, by studying the density of near extreme eigenvalues, also called the density of states, $\rho_{\rm DOS}(r,N)$, defined as \cite{SM2007,PS14} (see Fig. \ref{fig_illust})
\begin{eqnarray}\label{eq:def_rho_eigen}
\rho_{\rm DOS}(r,N)=\frac{1}{N-1} \sum_{\underset{i \ne i_{\rm max}}{i=1}}^{N} \langle \delta(\lambda_{\rm max}-\lambda_i-r) \rangle \;,
\end{eqnarray}
where $i_{\rm max}$ is such that $\lambda_{i_{\max}} = \lambda_{\rm max}$ and 
$\langle \cdots \rangle$ means an average taken with the weight in (\ref{jPDF}). It is normalized according to
\begin{eqnarray}
\int_0^{\infty}{\rm d}r\, \rho_{\rm DOS}(r,N)=1 \;.
\end{eqnarray}

Incidentally, we showed that the DOS is related to another interesting observable characterizing also the crowding to $\lmax$, namely   
the PDF of the gap between the two largest eigenvalues \cite{PS14}. Let us denote by $\lmax = \Lambda_{1,N} \geq \Lambda_{2,N} \geq \cdots \geq \Lambda_{N,N}$ and by $d_{1,N} = \Lambda_{1,N} - \Lambda_{2,N}$ the first gap (see Fig. \ref{fig_illust}). Its PDF is denoted by $p_{\rm GAP}(r,N)$, such that ${{\rm Pr.}} [d_{1,N} \in [r,r+\rmd r]] = p_{\rm GAP}(r,N) \rmd r$. It is then possible to show the following relation:
\begin{eqnarray}\label{id_gap_DOS}
p_{\rm GAP}(r,N) = (N-1) \rho_{\rm DOS}(-r,N) \;,
\end{eqnarray}  
and we refer the reader to Ref. \cite{PS14} for the derivation of this relation. 


In Ref. \cite{PS14}, we focused essentially on the case of GUE ($\beta = 2$) where we computed exactly these two quantities $\rho_{\rm DOS}(r,N)$ and $p_{\rm GAP}(r,N)$ both for finite $N$ and in the large $N$ limit, using a method based on (semi-classical) orthogonal polynomials, which were previously introduced in Ref. \cite{NM11}. Here, we generalize these results to the case of generic $\beta$, which we mainly study, for large $N$, using heuristic arguments, that are confirmed by numerical simulations (for $\beta =1, 2$ and $\beta=4$). Before presenting our results we discuss two applications of these two near extreme observables (\ref{eq:def_rho_eigen}, \ref{id_gap_DOS}).


\section{Two applications of $\rho_{\rm DOS}(r,N)$ and $p_{\rm GAP}(r,N)$}\label{sec:appli}

As we discuss it here, $\rho_{\rm DOS}(r,N)$ and $p_{\rm GAP}(r,N)$ play an important role in the minimization of quadratic forms on the sphere \cite{fyodorov_private} and in the study of a spherical mean-field spin-glass \cite{MG13} respectively.  

\subsection{Minimizing a quadratic form on the $N$-dimensional sphere}

Let us consider the problem which consists in minimizing a random quadratic form on the $N$-dimensional sphere $S_N$ :
\begin{eqnarray}\label{def_spherical}
H[\vec{s}\,]=-\frac{1}{2}\sum_{i,j=1}^N J_{i,j} s_i s_j \,, \hspace{0.5cm} \vec{s}\,^2=\sum_{i=1}^N s_i^2 = N \, ,
\end{eqnarray}
where ${\bf J}$ is a matrix belonging to the Gaussian $\beta$-ensemble (with $\beta = 1,2$ being the more natural values in this case) -- and here we choose the matrix elements $J_{i,j}$ of zero mean and variance of order ${\cal O}(1/N)$ \footnote{note that the joint PDF (\ref{jPDF}) corresponds instead to matrix elements of order ${\cal O}(1)$.}.
To take into account the spherical constraint (\ref{def_spherical}) we introduce a Lagrange multiplier $z$ such that we have to minimize
\begin{eqnarray}
\tilde{H}[\vec{s}\,,z]=-\frac{1}{2}\sum_{i,j=1}^N J_{i,j} s_i s_j + z\left(\vec{s}\,^2-N\right)
\end{eqnarray}
with respect to $\vec{s}$ and $z$. It is then straightforward to show that
\begin{eqnarray}
\underset{\vec{s}\,,z}{\min} \; \tilde{H}[\vec{s},z]=\tilde{H}[\vec{s}_{\max},z_{\max}]=-N \frac{\lmax}{2} \,, 
\end{eqnarray}
where $\vec{s}_{\max}$ and $z_{\max}$ are such that 
\begin{eqnarray}
\begin{cases}
{\bf J} \, \vec{s}_{\max} = \lmax \vec{s}_{\max} \;, \\
z_{\max} = \dfrac{\lmax}{2} \;.
\end{cases}
\end{eqnarray}
If we now look at the Hessian matrix ${\delta^2 \tilde{H}}/{\delta s_i \delta s_j}$ evaluated at the minimum $\vec{s}_{\max},z_{\max}$, one can show that its spectrum is given by
\begin{eqnarray}
{\rm Sp} \left[ \frac{\delta^2 \tilde{H}}{\delta s_i \delta s_j} \Big|_{\vec{s}_{\max},z_{\max}}\right] = \{0, \lmax-\lambda_1, \lmax-\lambda_2,\cdots, \lmax-\lambda_N\} \;.
\end{eqnarray}
Hence the DOS, $\rho_{\rm DOS}(r,N)$ in Eq. (\ref{eq:def_rho_eigen}) is the average density of eigenvalues of the Hessian evaluated at the minimum (except the trivial eigenvalue $0$). It is thus natural to expect that the DOS plays an important role in the relaxational dynamics of a system driven by such a quadratic form (\ref{def_spherical}), which corresponds precisely to the fully connected $p$-spin spherical spin-glass model with $p=2$ (where the $s_i$'s correspond to spin variables coupled to each other via the matrix ${\bf J}$) \cite{cugliandolo_dean,kurchan_laloux,us_spherical}. 


\subsection{Overlap distribution in the fully connected $p=2$-spherical spin glass model}

In Ref. \cite{MG13}, the authors studied the equilibrium properties of the $p=2$-spherical spin glass model (or spherical Sherrington-Kirkpatrick model) described by the partition function associated to the above Hamiltonian (\ref{def_spherical}) 
\begin{eqnarray}
Z = \int_{-\infty}^\infty \rmd s_1 \cdots \int_{-\infty}^\infty \rmd s_N e^{-\frac{1}{2T} \, \sum_{i,j} J_{i,j} s_i s_j} \,\delta\left(\sum_{i=1}^N s_i^2 - N\right) \;,
\end{eqnarray}
where the $\delta$ function ensures the spherical constraint (\ref{def_spherical}). They focused on the overlap $Q$, which is the order parameter characterizing the spin-glass order, defined by
\begin{eqnarray}
Q = \sum_{i=1}^N s_i^{(1)} s_i^{(2)} \;,
\end{eqnarray}
where $s_{i}^{(1)}$ and $s_i^{(2)}$ represent the spins at site $i$ in two distinct equilibrium configurations with the same realization of the couplings $J_{i,j}$. In Ref. \cite{MG13} it was shown that, for fixed couplings $J_{i,j}$, the distribution of $Q$ (with respect to thermal fluctuations) is related, at low temperature $T$, to the first gap between the two largest eigenvalues $\Lambda_{1,N}$ and $\Lambda_{2,N}$ of the matrix ${\bf J}$
\begin{eqnarray}\label{overlap}
P(Q,N) \sim \frac{1}{N} \tilde P\left(q=\frac{Q}{N}\right) \;, \; \tilde P(q) = e^{-\frac{N}{2T} (\Lambda_{1,N} - \Lambda_{2,N})(1-q^2)} \;.
\end{eqnarray}
Hence we see from (\ref{overlap}) that the full distribution of the overlap in this model is directly related to the PDF of the first gap $d_{1,N} = \Lambda_{1,N} - \Lambda_{2,N}$ of Gaussian random matrices $J_{i,j}$, the natural ensemble being here GOE ($\beta = 1$). Indeed, after averaging the distribution $P(Q,N)$ in (\ref{overlap}) over the random couplings $J_{i,j}$, one obtains nothing else but the Laplace transform (with Laplace parameter $\propto (1-q^2)/T$) of the PDF of the first gap, $p_{\rm GAP}(r,N)$, studied here [see Eq.~(\ref{id_gap_DOS})]. 

\section{Two different scaling regimes: bulk and edge}

Computing the DOS $\rho_{\rm DOS}(r,N)$ for finite $N$ is, for a generic value of $\beta$ in (\ref{jPDF}), a very challenging task. This could be done for the special case $\beta = 2$ in Ref. \cite{PS14} using the method of orthogonal polynomials, which unfortunately can not be extended to other values of $\beta$. In spite of this difficulty, the main features of  $\rho_{\rm DOS}(r,N)$, for large $N$, can be characterized by means of heuristic arguments.  

For this purpose, it is first useful to recall that the fluctuations of the eigenvalues are characterized by two different scales depending on their location in the spectrum: (i) in the bulk for $\lambda_i/\sqrt{N} = {\cal O}(1)$ and $|\lambda_i| < \sqrt{2N}$ and (ii) at the edge where $|\lambda_i \pm \sqrt{2N}| = {\cal O}(N^{-1/6})$. The existence of these two scales manifests itself in various observables associated to the eigenvalues of the Gaussian $\beta$-ensemble (\ref{jPDF}), including their mean density defined as:
\begin{eqnarray}\label{def_rho}
\rho(\lambda,N) = \frac{1}{N} \sum^N_{i=1} \langle \delta(\lambda_i - \lambda) \rangle \;.
\end{eqnarray} 
One has obviously $\rho(\lambda,N) = \rho(-\lambda,N)$ and one can further show that, for large $N$, it exhibits two distinct regimes (for $\lambda > 0$)~\cite{Meh91,For10,BB91,For93}
\begin{eqnarray}\label{scaling_density}
\rho(\lambda,N) \sim
\begin{cases}
\dfrac{1}{\sqrt{N}} \rho_{\rm bulk}\left(\dfrac{\lambda}{\sqrt{N}} \right) \;,\; \lambda < \sqrt{2N} \; \;\& \; &\lambda = {\cal O}(\sqrt{N}) \;, \\
\\
\dfrac{\sqrt{2}}{N^{5/6}} \rho_{\rm edge}\left( (\lambda-\sqrt{2N})\sqrt{2}N^{1/6} \right) \;, &|\lambda - \sqrt{2N}| = {\cal O}(N^{-1/6}) \;.
\end{cases}
\end{eqnarray}
In Eq. (\ref{scaling_density}), $\rho_{\rm bulk}(x)$ is the Wigner semi-circle \cite{Meh91,For10}:
\begin{eqnarray}\label{eq:semi_circle}
\rho_{\rm bulk}(x) = \rho_W(x) = \frac{1}{\pi} \sqrt{2-x^2} \;,
\end{eqnarray}
independently of $\beta$, while $\rho_{\rm edge}(x)$ is given by \cite{BB91,For93,For06},
\begin{eqnarray}
\rho_{\rm edge}(x) \sim
\begin{cases}
 [{\rm Ai}'(x)]^2 - x {\rm Ai}^2(x)+\frac12 {\rm Ai}(x)\left(1-\int_x^\infty \rmd t {\rm Ai}(t) \right)  \;, \; \beta=1 \;,\hspace{0.4cm}\\
 \\
 [{\rm Ai}'(x)]^2 - x {\rm Ai}^2(x)  \;, \; \hspace{4.6cm}\beta=2 \;,\\
 \\
 \kappa^{-1/2}\Big([{\rm Ai}'(\kappa x)]^2 - \kappa x {\rm Ai}^2(\kappa x) -\frac12 {\rm Ai}(\kappa x)\int_{\kappa x}^\infty \rmd t {\rm Ai}(t) \Big) \;, \;\\
\hspace{8.05cm}\; \beta=4 \;,\\
\end{cases}
\end{eqnarray}
where $\kappa = 2^{2/3}$.
Its asymptotic behaviors are given by 
\begin{eqnarray}\label{asympt_edge}
\begin{cases}
&\rho_{\rm edge}(x) \sim\dfrac{1}{\pi}\sqrt{-x} \;, \hspace{1.4cm}\; x \to -\infty \;,\\
& \\
& \ln(\rho_{\rm edge}(x) ) \sim -\frac{2\beta}{3}x^{3/2}\;, \hspace{0.55cm}\; x \to \infty \;.
\end{cases}
\end{eqnarray}

Interestingly, one can check that these two regimes for $\rho(\lambda,N)$, the ``bulk'' one and the ``edge'' one in Eq. (\ref{scaling_density}), 
perfectly match when $\lambda$ approaches the value $\sqrt{2N}$ from below. Indeed, when $\lambda \to \sqrt{2N}$ from below, $\rho(\lambda,N)$ can be replaced by the Wigner semi-circle (\ref{eq:semi_circle}), which gives:  
\begin{eqnarray}\label{eq:matching_left}
\rho(\lambda,N) \sim \frac{2^{3/4}}{\pi} N^{-3/4} \left({\sqrt{2N} - \lambda}\right)^{1/2} \;, \; \lambda \to \sqrt{2N}^{\,-} \;.
\end{eqnarray}
This behavior (\ref{eq:matching_left}) coincides with the left tail of the scaling function $\rho_{\rm edge}(x)$ in Eq. (\ref{asympt_edge}). Indeed, when the deviation from $\sqrt{2N}$ is large, $\sqrt{2N} - \lambda \sim {\cal O}(\sqrt{N})$, we can substitute in the second line of Eq. (\ref{scaling_density}) the left tail asymptotic behavior of $\rho_{\rm edge}(x)$ in (\ref{asympt_edge}), which gives
\begin{eqnarray}\label{eq:matching_right}
\rho(\lambda,N) \sim \sqrt{2}N^{-5/6} \frac{1}{\pi} \left(\sqrt{2}N^{1/6}(\sqrt{2N}-\lambda) \right)^{1/2} \;, \; \lambda \to \sqrt{2N}^{\,-} \;,
\end{eqnarray} 
which after a trivial rearrangement coincides with Eq. (\ref{eq:matching_left}). Note also that the right tail of $\rho_{\rm edge}(x)$ in Eq. (\ref{asympt_edge}) matches, as it should, with the right tail of the TW distribution for $\beta$-ensemble \cite{For10}.

\begin{figure}[h!]
\begin{center}
\rotatebox{-90}{\resizebox{!}{12.5cm}{\includegraphics{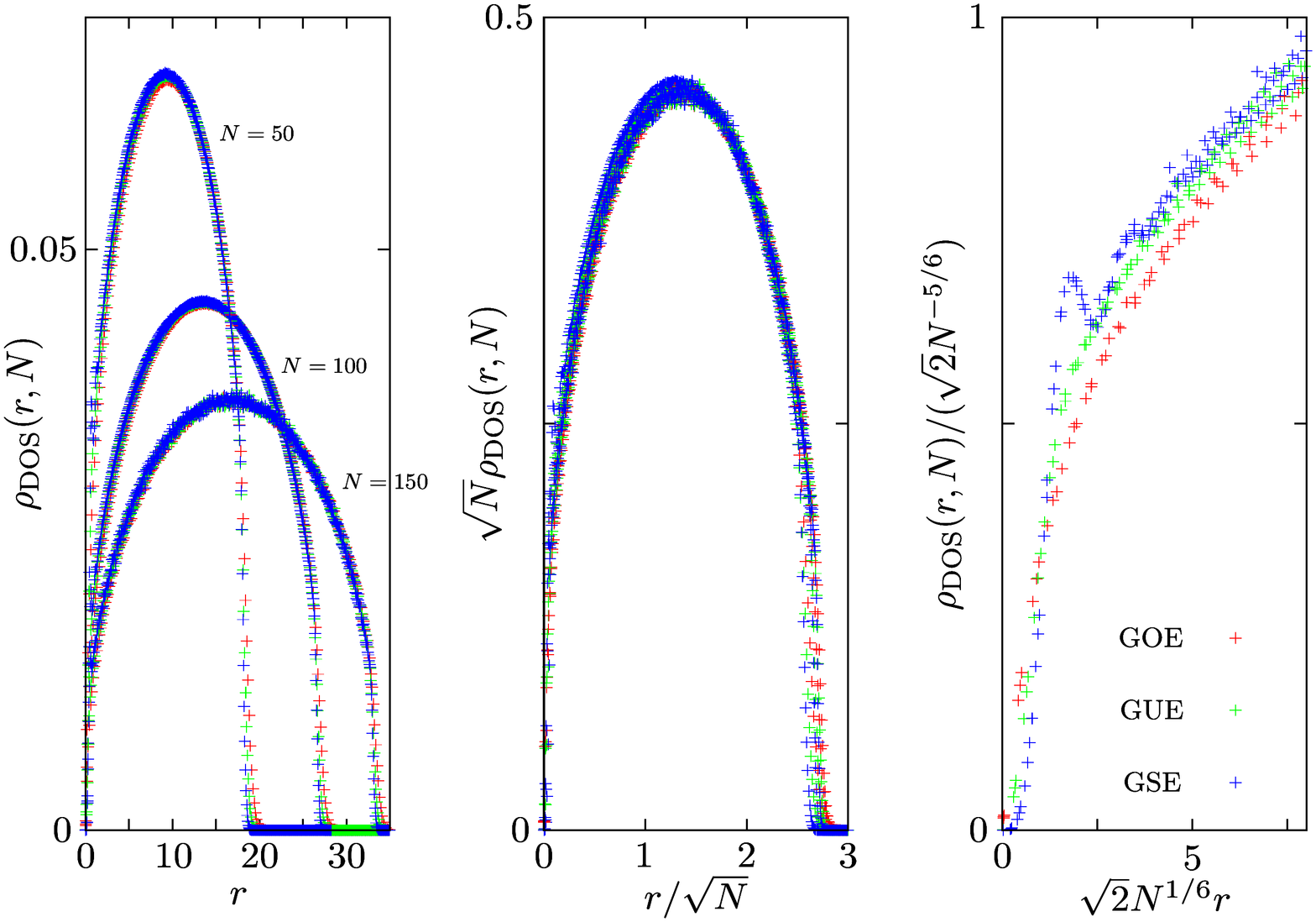}}}
\vspace{-0.2cm}
\caption{The different scaling regimes for $\rho_{\rm DOS}$ as described in Eq. (\ref{DOS_main}). {\bf Left panel:} Plot of $\rho_{\rm DOS}(r,N)$ as a function of $r$ for three different values of $N$ and three different values: $\beta=1$ (red), $\beta = 2$ (green) and $\beta = 4$ (blue). {\bf Central panel:} (bulk regime) Plot of the same quantities (after rescaling)  $\sqrt{N}\rho_{\rm DOS}(r,N)$ as a function of $r/\sqrt{N}$ and for different value of $\beta$ with the same color code as the one used in the left panel. {\bf Right panel:} (edge regime) Plot of the same quantities (after yet another rescaling) $\rho_{\rm DOS}(r,N)/(\sqrt{2}N^{-5/6})$ as a function of $\sqrt{2}N^{1/6} r$.}
\label{rho_dos_scaling}
\end{center} 
\end{figure}

Similarly to the density (\ref{scaling_density}), one expects that, for large $N$, $\rho_{\rm DOS}(r,N)$ exhibits two different scaling regimes (see figure \ref{rho_dos_scaling}): (i) a bulk regime, where $r \propto \sqrt{N}$ and (ii) an edge regime where $r = {\cal O}(N^{-1/6})$. They can thus be summarized as follows
\begin{eqnarray}\label{DOS_main}
\rho_{\rm DOS}(r,N) \sim
\begin{cases}
\dfrac{1}{\sqrt{N}} \tilde \rho_{\rm bulk} \left( \dfrac{r}{\sqrt{N}} \right) \;, &c\,\sqrt{N} < r < 2\sqrt{2N} \;, \\
& \\
\sqrt{2} N^{-5/6} \tilde \rho_{\rm edge}\left( r\sqrt{2}N^{1/6} \right) \;, \; &r = {\cal O}(N^{-1/6}) \;,
\end{cases}
\end{eqnarray}
for some real $c< 2\sqrt{2}$ and where $\tilde \rho_{\rm bulk}(x)$ and $\tilde \rho_{\rm edge}(\tilde r)$ are two different scaling functions. 

Let us first investigate the bulk regime. For $r \propto \sqrt{N}$ in (\ref{eq:def_rho_eigen}), one expects that $\rho_{\rm DOS}(r,N)$ is insensitive to the fluctuations of $\lambda_{\max}$ which are of order ${\cal O}(N^{-1/6})$ around the value $\sqrt{2N}$. Therefore, in Eq. (\ref{eq:def_rho_eigen}) the PDF of $\lmax$ can be simply replaced by a delta function $\delta(\lmax - \sqrt{2N})$. It thus follows that $\rho_{\rm DOS}(r,N) \approx \rho(\sqrt{2N} - r,N)$ where $\rho(\lambda,N)$ is simply the density of eigenvalues in (\ref{def_rho}). Therefore, from (\ref{scaling_density}) together with (\ref{eq:semi_circle}) one expects that $\rho_{\rm DOS}(r,N)$ takes the scaling form given in the first line of Eq.~(\ref{DOS_main}) where $\tilde \rho_{\rm bulk}(x)$ is a shifted Wigner semi-circle: 
\begin{eqnarray}\label{shifted_wigner}
\tilde \rho_{\rm bulk}(x) = \rho_W(\sqrt{2}-x)=\frac{1}{\pi} \sqrt{x(2\sqrt{2}-x)} \;,
\end{eqnarray}
independently of $\beta$. 

On the other hand, in the edge regime when $r \sim {\cal O}(N^{-1/6})$, the DOS will be sensitive to the fluctuations of $\lmax$ and we expect that $\tilde \rho_{\rm edge}(\tilde r)$ is a non trivial function, as it was shown to be the case for $\beta = 2$ \cite{PS14} (see also Eq.~(\ref{rho_final_simplif}) below). Although we can not compute explicitly this scaling function $\tilde \rho_{\rm edge}(\tilde r)$ for other values of $\beta \neq 2$ we can extract its asymptotic behaviors for both small and large arguments, which we discuss in the next section.

In Ref. \cite{PS14} these results were obtained by exact analytical calculations for $\beta = 2$. Here we have performed numerical simulations for different values of $\beta = 1, 2$ and $4$ using the tridiagonal random matrices representation \cite{DE02}. In Fig. \ref{rho_dos_scaling} we show a plot of $\rho_{\rm DOS}(r,N)$ as a function of $r$ and different values of $N$ that corroborates the scaling forms in Eq. (\ref{DOS_main}) for three different values of $\beta =1$ (red), $\beta = 2$ (green) and $\beta = 4$ (blue). The central panel indicates that, in the bulk regime, $\rho_{\rm DOS}(r,N)$, correctly rescaled, converges to a shifted Wigner law, independently of $\beta$. This is in full agreement with the scaling form in the first line of Eq. (\ref{DOS_main}). Finally, the right panel shows a plot of $\rho_{\rm DOS}(r,N)$ for small $r$, which is in a good agreement with the scaling form given in the second line of Eq. (\ref{DOS_main}). It also indicates that the limiting function $\tilde \rho_{\rm edge}(\tilde r)$ depends explicitly on $\beta$. 

We conclude this section by mentioning that the typical fluctuations of the first gap are naturally expected to scale as $d_{1,N} = \Lambda_{1,N} - \Lambda_{2,N} \sim {\cal O}(N^{-1/6})$, as the fluctuations of $\Lambda_{1,N} = \lmax$ and $\Lambda_{2,N}$ around $\sqrt{2N}$
are also of order ${\cal O}(N^{-1/6})$ \cite{TW94a,TW96}. Hence, for large $N$ we expect that $p_{\rm GAP}(r,N)$ takes the scaling form:
\begin{eqnarray}\label{scaling_gap}
p_{\rm GAP}(r,N) = \sqrt{2} N^{1/6} \tilde p_{\rm typ}\left( r\sqrt{2}N^{1/6} \right) \;,
\end{eqnarray}
where the factor $\sqrt{2}$ in the argument has been chosen here for convenience, according to the choice made for $\rho_{\rm DOS}(r,N)$ in the second line of Eq. (\ref{DOS_main}). This scaling form (\ref{scaling_gap}) together with the scaling function $\tilde p_{\rm typ}(\tilde r)$ were obtained exactly for $\beta = 2$. It is natural to expect that this scaling form (\ref{scaling_gap}) holds for any value of $\beta > 0$, with a $\beta$-dependent scaling function $\tilde p_{\rm typ}(\tilde r)$, and below we give some heuristic arguments to obtain the asymptotic behaviors of the scaling function $\tilde p_{\rm typ}(\tilde r)$.

\section{Asymptotic behaviors of the scaling functions $\tilde \rho_{\rm edge}(\tilde r)$ and~$\tilde p_{\rm typ}(\tilde r)$}

We first begin to analyze the small $\tilde r$ behavior of $\tilde \rho_{\rm edge}(\tilde r)$. From the definition of the DOS in Eq. (\ref{eq:def_rho_eigen}) it is clear that its small $\tilde r$ behavior is directly related to the probability that two eigenvalues -- namely the first one and the second one -- become extremely close to each other. From the joint PDF of the eigenvalues (\ref{jPDF}), this probability vanishes as ${\tilde r}^\beta$ as a consequence of the short distance repulsion between eigenvalues, which comes from the Vandermonde term $\prod_{i,j}|\lambda_i - \lambda_j|^\beta$ in the joint PDF (\ref{jPDF}). One thus expects that $\tilde \rho_{\rm edge}(\tilde r) \sim a_\beta\, \tilde r^\beta$, with an a priori unknown constant $a_\beta$. 

On the other hand, it is reasonable to assume that there is a smooth matching between the edge region and the bulk region described by the shifted Wigner semi-circle law (\ref{shifted_wigner}), as it is the case for the density of eigenvalues [see the discussion between Eqs. (\ref{eq:matching_left}) and (\ref{eq:matching_right})]. This means that the large $\tilde r$ behavior of $\tilde \rho_{\rm edge}(\tilde r)$ has to coincide with the small argument of the shifted Wigner semi circle (\ref{shifted_wigner}), which does not depend on $\beta$. Hence one deduces that $\tilde \rho_{\rm edge}(\tilde r) \sim {\sqrt{\tilde r}}/{\pi}$, for all $\beta$. The asymptotic behaviors of the DOS in the edge regime can thus be summarized as follows:
\begin{eqnarray}
\tilde \rho_{\rm edge}(\tilde r) \sim
\begin{cases}\label{asympt_tilderho_beta}
&a_\beta \, \tilde r^\beta  + o(\tilde r^{\beta}) \;, \; \tilde r \to 0\,,\\
& \\
&\dfrac{\sqrt{\tilde r}}{\pi} + {o}(\tilde r^{1/2}) \;, \; \tilde r \to \infty.\\
\end{cases}
\end{eqnarray}

\begin{figure}[h!]
\begin{center}
\rotatebox{-90}{\resizebox{8cm}{!}{\includegraphics{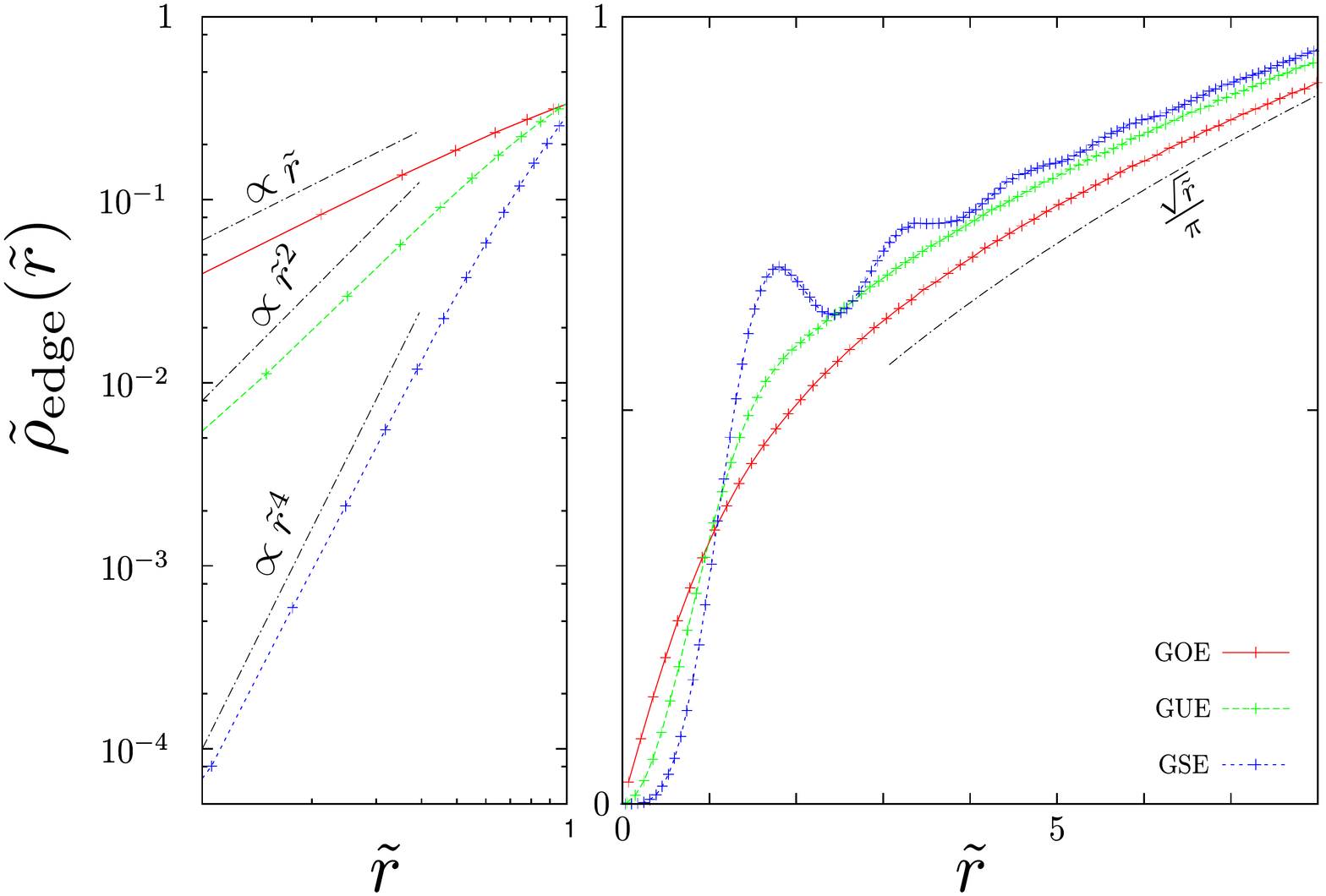}}}
\vspace{-0.2cm}
\caption{Plot of $\tilde{\rho}_{\rm edge}(\tilde{r})$ for three different values of $\beta$ ($\beta = 1$ in red, $\beta = 2$ in green and $\beta = 4$ in blue). {\bf Left panel:} the data are plotted on a log-log scale and they exhibit a small $\tilde r$ behavior compatible with our heuristic arguments in (\ref{asympt_tilderho_beta}). {\bf Right panel:} plot of $\tilde \rho_{\rm edge}(\tilde r)$ in terms of $\tilde r$ with the asymptotic behavior for large $\tilde r$ compatible with our heuristic arguments (\ref{asympt_tilderho_beta}).} 
\label{rhoedge_betacase}
\end{center}
\end{figure}

These behaviors (\ref{asympt_tilderho_beta}) have been confirmed, for $\beta = 2$, by an exact calculation of $\tilde \rho_{\rm edge}(\tilde r)$~\cite{PS14}:
\begin{eqnarray}\label{rho_final_simplif}
\tilde \rho_{\rm edge}(\tilde r) 
= \frac{2^{1/3}}{\pi} \int_{-\infty}^\infty \left[ \tilde f(\tilde r,u)^2 - \left( \int_x^\infty q(u) \tilde f(\tilde r,u) \rmd u  \right)^2 \right] {\cal F}_2(x) \, \rmd x \;,
\end{eqnarray}
where ${\cal F}_2(x)$ is the Tracy-Widom distribution associated to GUE \cite{TW94a} and $q(x)$ is the Hastings-Mc Leod solution of the Painlev\'e II equation
\begin{eqnarray}
{\cal F}_2(x) = \exp{\left[-\int_x^\infty (u-x) q^2(u) \rmd u\right]} \;, \hspace{0.5cm}
\left\{
     \begin{array}{lr}
			q''(x)=2q^3(x)+x \, q(x) \;,\\
			\\
			q(x)\sim {\rm Ai}(x) \textrm{ for } x \to \infty \;,
     \end{array}
\right.
\end{eqnarray}
while $\tilde f(\tilde r,x)$ satisfies 
\begin{eqnarray}\label{schrod_text}
\left\{
     \begin{array}{lr}
\partial_x^2 \tilde f(\tilde r,x) - [x + 2 q^2(x)] \tilde f(\tilde r,x) = - \tilde r \tilde f(\tilde r,x) \;, \\
\\
 \tilde f(\tilde r,x) \underset{x \to \infty}{\sim} 2^{-1/6} \sqrt{\pi} {\rm Ai}(x-\tilde r) \;.
     \end{array}
\right.
\end{eqnarray}
It was further shown in \cite{PS14} that $\tilde f(\tilde r,x)$ can be expressed in terms of the Lax pair associated to the Painlev\'e XXXIV equation. From this exact expression (\ref{rho_final_simplif}) one can derive the asymptotic behaviors announced in (\ref{asympt_tilderho_beta}) and compute explicitly the amplitude $a_2 = 1/2$~\cite{PS14}.  

In Fig. \ref{rhoedge_betacase} we show a plot of $\tilde \rho_{\rm edge}(\tilde r)$ as a function of $\tilde r$ for three different values of $\beta = 1,2$ and $4$ exhibiting the small $\tilde r$ behavior (in the left panel) and the large $\tilde r$ behavior (in the right panel). In each case, $\tilde \rho_{\rm edge}(\tilde r)$ was computed by sampling $2. 10^7$ independent random matrices of size $100\times100$. Both panels show a good agreement with our predictions in (\ref{asympt_tilderho_beta}). We also notice on the right panel of Fig. \ref{rhoedge_betacase} that $\tilde \rho_{\rm edge}(\tilde r)$ exhibits clear oscillations for $\beta = 4$, while the curves for $\beta = 1$ and $2$ are much smoother.

We now turn to the small $\tilde r$ asymptotic behavior of the gap distribution $\tilde p_{\rm typ}(\tilde r)$, which can be argued to coincide with the small $\tilde r$ behavior of the DOS, $\tilde \rho_{\rm edge}(\tilde r)$. Indeed, for small $\tilde{r}$, the contribution to $\tilde{\rho}_{\rm edge}(\tilde{r})$ comes only from the gap between the two largest eigenvalues because the others (the third, the fourth etc. eigenvalues) are too far, as a consequence of the short range repulsion coming from the Vandermonde term in Eq. (\ref{jPDF}). Hence one expects that $\tilde p_{\rm typ}(\tilde r) \sim a_\beta\, \tilde r^\beta$ with
the same amplitude $a_\beta$ as above in (\ref{asympt_tilderho_beta}).  

On the other hand, the large $\tilde r$ behavior of $\tilde p_{\rm typ}(\tilde r)$ can be obtained, to leading order, by assuming that for large separation $\tilde r$ the two first eigenvalues $\Lambda_{2,N}$ and $\Lambda_{1,N} = \lmax$ become statistically independent. Hence one expects:
\begin{eqnarray}\label{heuristic}
p_{\rm GAP}(r,N) &=& \int_{-\infty}^\infty {\rm Pr.} [\Lambda_{2,N} = \lambda, \lmax = \lambda+ r] \rmd\lambda \nn\\
&\underset{r \to \infty}{\sim}& \int_{-\infty}^\infty {\rm Pr.} [\Lambda_{2,N} = \lambda] {\rm Pr.} [\lmax = \lambda+ r] \rmd\lambda \\
&\underset{r \to \infty}{\sim}& {\rm Pr.} [\lmax = r] \nn\;.
\end{eqnarray}
Therefore, we expect that right tail of $\tilde p_{\rm typ}(\tilde r)$ coincides with the right tail  
of the TW distribution which goes for large argument as $\ln {\cal F}'_\beta(x) \sim -\frac{2\beta}{3}x^{3/2}$. To summarize, we obtain the asymptotic behaviors of $\tilde p_{\rm typ}(\tilde r)$ as:
\begin{eqnarray}\label{asympt_tildegap_beta}
\begin{cases}
\tilde p_{\rm typ}(\tilde r) \sim a_{\beta} \, \tilde r^\beta  + o(\tilde r^\beta) & \tilde r \to 0 \\
\\
\ln \tilde p_{\rm typ}(\tilde r) \sim -\dfrac{2\beta}{3}\tilde r^{3/2} + o(\tilde r^{3/2})  \;, \; & \tilde r \to +\infty \;,
\end{cases}
\end{eqnarray} 
while the computation of $a_{\beta}$ as well as the subleading corrections, both for small and large arguments, for any $\beta$ remains challenging. 

In Ref. \cite{PS14}, we obtained an exact expression of $\tilde p_{\rm typ}(\tilde r)$ for the special case $\beta = 2$ as
\begin{eqnarray}\label{gap_intro_simplif}
\tilde p_{\rm typ}(\tilde r) = \frac{2^{1/3}}{\pi} \int_{-\infty}^\infty \left[\tilde f^2(-\tilde r,x) - \left( \int_x^\infty q(u) \tilde f(-\tilde r,u) \rmd u  \right)^2 \right] {\cal F}_2(x) \, \rmd x \;.
\end{eqnarray}
From this exact formula (\ref{gap_intro_simplif}) we could not only check the above asymptotic behaviors (\ref{asympt_tildegap_beta}) but also obtain the sub-leading terms as:
\begin{equation}\label{asympt_tildegap}
\tilde p_{\rm typ}(\tilde r) = 
\begin{cases}
\frac{1}{2} \tilde r^2 + a_4 \tilde r^4 + {\cal O}(\tilde r^6) \\
\\
A \exp{\left(-\dfrac{4}{3}\tilde r^{3/2} + \dfrac{8}{3}\sqrt{2} \, \tilde r^{3/4}\right)}{\tilde r^{-\frac{21}{32}}} \left(1 - \dfrac{1405 \sqrt{2}}{1536} \tilde r^{-3/4} + {\cal O}(\tilde r^{-3/2})\right)\,, 
\end{cases}
\end{equation} 
where the first and second lines correspond respectively to the small and large $\tilde r$ behaviors. In (\ref{asympt_tildegap}),  
the amplitude $A$ is given by $A = 2^{-91/48} e^{\zeta'(-1)}/\sqrt{\pi}$, where $\zeta'(x)$ is the derivative of the Riemann zeta function, while $a_4$ can be expressed in terms of integrals involving $q(x)$ with the result $a_4\sim -0.393575...$. It should be noticed that a different expression, somehow more complicated, of the PDF of the first gap $\tilde p_{\rm typ}(\tilde r)$ had been obtained previously in Ref.~\cite{WBF13}, involving also Painlev\'e transcendents. It remains an open question to show that these two expressions do coincide.

\begin{figure}
\begin{center}
\resizebox{!}{8.5cm}{\includegraphics{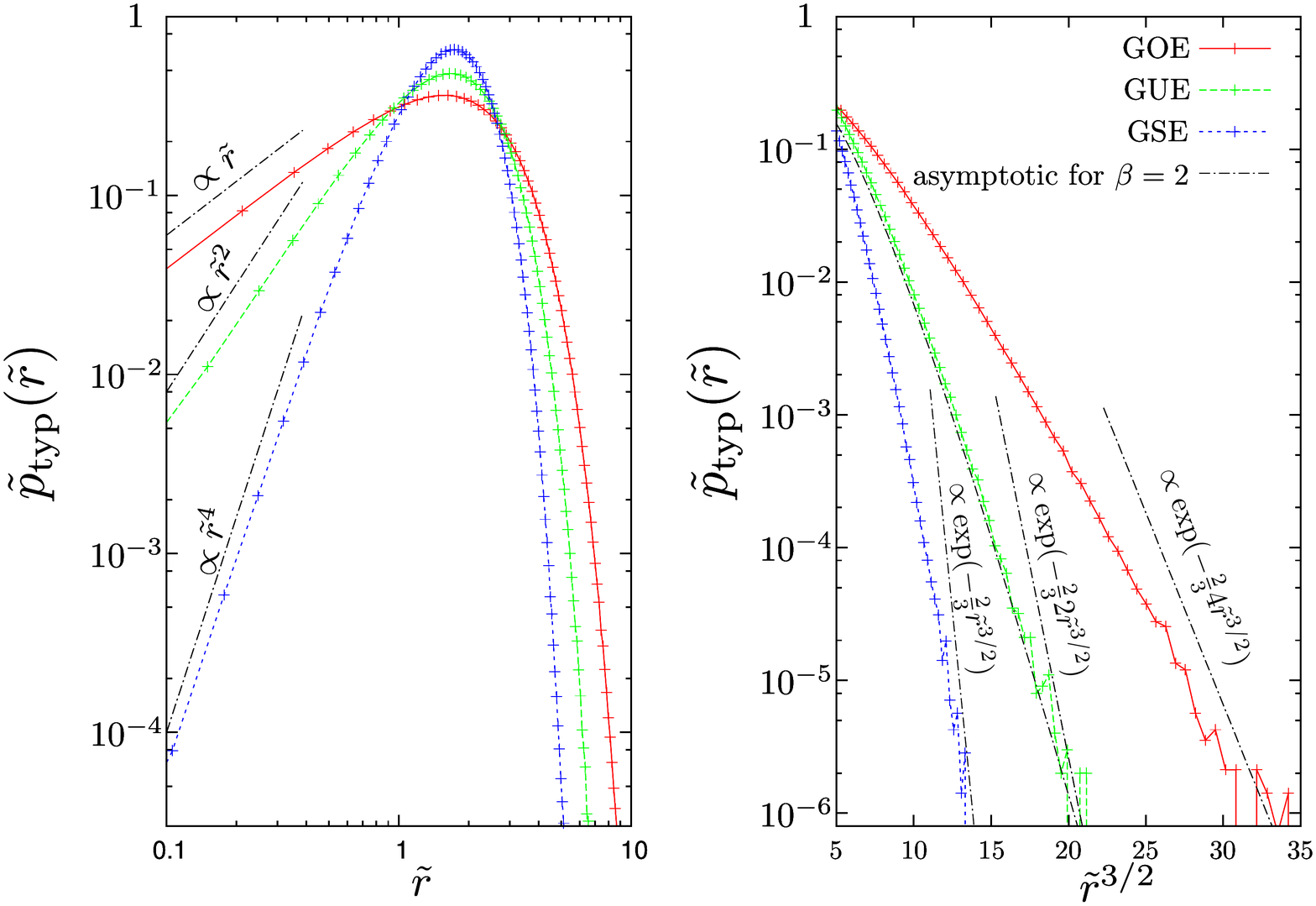}}
\vspace{-0.2cm}
\caption{Comparison of $\tilde{p}_{\rm typ}(\tilde{r})$ for the different $\beta$-ensemble with $2.10^7$ realizations for a $100 \times 100$ matrix for each case. ($\beta=1$ in red, $\beta=2$ in green and $\beta=4$ in blue) {\bf Left panel:} log-log plot of $\tilde{p}_{\rm typ}(\tilde{r})$ with the asymptotic behavior for small $\tilde r$. {\bf Right panel:} log plot of $\tilde{p}_{\rm typ}$ in terms of $\tilde{r}^{3/2}$ with the asymptotic behavior for large $\tilde r$. We also plot the asymptotic behavior for large $\tilde r$ obtained in the $\beta=2$ case Eq. (\ref{asympt_tildegap}).}
\label{ptyp_betacase}
\end{center}
\end{figure}

\section{The density of states and the gap for a fixed value of $\lmax$}

\begin{figure}[hh]
\begin{center}
\rotatebox{-90}{\resizebox{8cm}{!}{\includegraphics{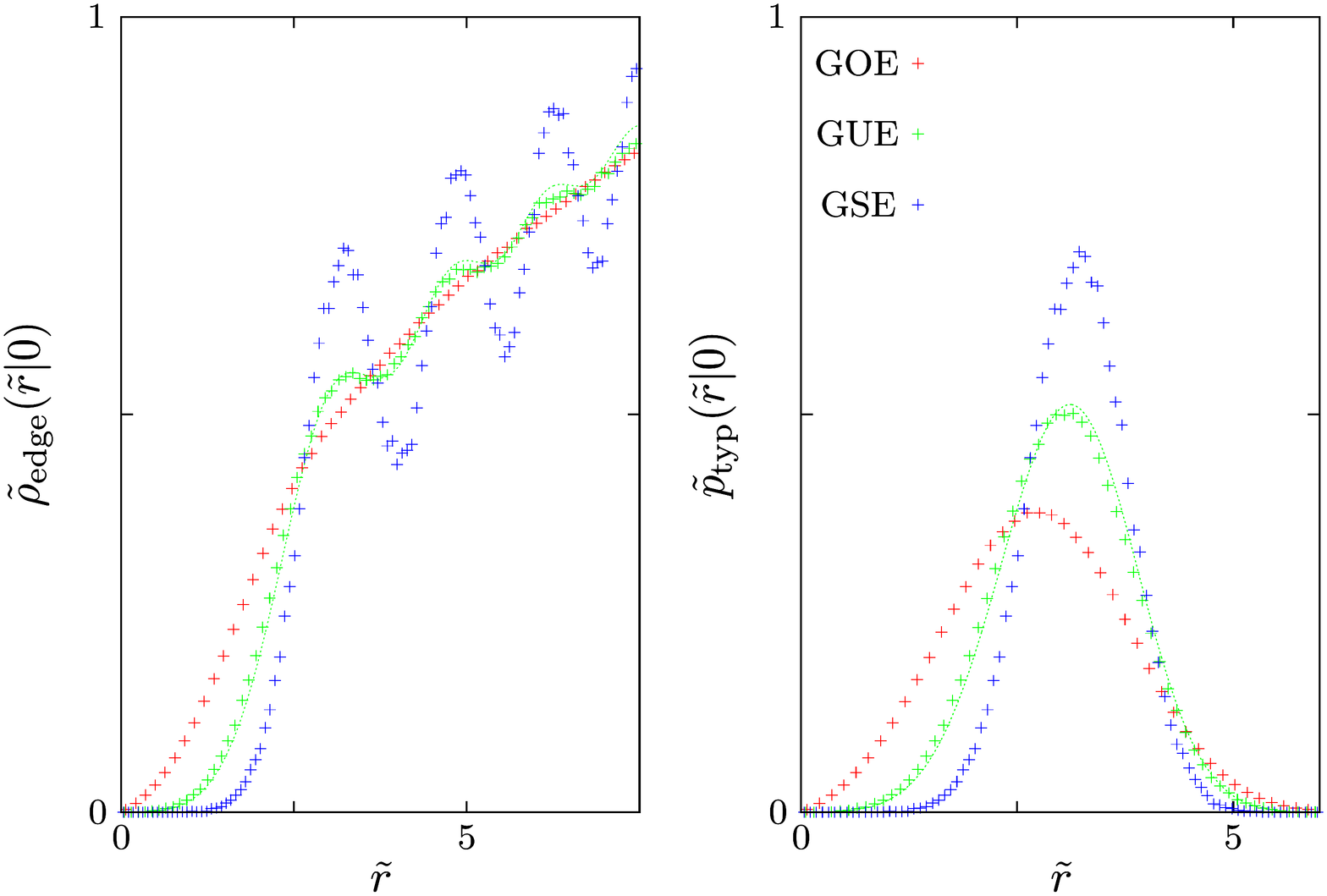}}}
\vspace{-0.2cm}
\caption{
Comparison of $\tilde{\rho}_{\rm edge}(\tilde{r}|0)$ {\bf (left panel)} and $\tilde{p}_{\rm typ}(\tilde{r}|0)$ {\bf (right panel)} for three different values of $\beta$ ($\beta=1$ in red, $\beta=2$ in green and $\beta=4$ in blue). We have performed $2.10^7$ realizations of a $200 \times 200$ matrix. For each case, we keep only events when $|\lambda_{\max}-\sqrt{2N}| < 0.1$ (dots). We also plot in solid line the exact result for $\beta=2$ (\ref{result_rho_edge_contraint}) and (\ref{result_p_typ_contraint}) obtained in \cite{PS14} (using the numerical values of $q(x)$ and ${\cal F}_2(x)$ from Ref. \cite{Praehofer_webpage}).}
\label{rhoedge_pgap_contraint}
\end{center}
\end{figure}

Up to now, we have studied the DOS $\rho_{\rm DOS}(r,N)$ and the PDF of the gap $p_{\rm GAP}(r,N)$ averaged over the value of $\lmax$. It is also interesting to look at these quantities for fixed value of $\lmax = y$, the corresponding quantities being denoted as $\rho_{\rm DOS}(r|y,N)$ and $p_{\rm GAP}(r|y,N)$. We naturally expect the scaling forms, valid for all $\beta >0$: 
\begin{eqnarray}
&\rho_{\rm DOS}(r|y,N)=N^{-5/6}\sqrt{2}\hspace{0.1cm}\tilde{\rho}_{\rm edge}(\sqrt{2}N^{1/6}r|\sqrt{2}N^{1/6}(y-\sqrt{2N}))\, ,\\
&p_{\rm GAP}(r|y,N)=N^{1/6}\sqrt{2}\hspace{0.1cm}\tilde{p}_{\rm typ}(\sqrt{2}N^{1/6}r|\sqrt{2}N^{1/6}(y-\sqrt{2N}))\, .
\end{eqnarray}
From the results of Ref. \cite{PS14} one can compute explicitly these scaling functions as:
\begin{eqnarray}
\tilde{\rho}_{\rm edge}(\tilde{r}|x)=\frac{2^{1/3}}{\pi}{\cal F}_2(x)&\Bigg[&R(x)\Big(\left(\tilde{r}+\frac{R(x)}{q^2(x)}\right)\tilde{f}^2(\tilde{r},x)-2\frac{q'(x)}{q(x)}\tilde{f}(\tilde{r},x)\tilde{g}(\tilde{r},x) \nonumber \\
&&+ \left(1+\frac{q^2(x)}{\tilde{r}}\right)\tilde{g}(\tilde{r},x)^2\Big)-\frac{q^2(x)}{\tilde{r}^2}\tilde{g}^2(\tilde{r},x)\Bigg] \,,
\label{result_rho_edge_contraint}
\end{eqnarray}
with $R(x)=\int_x^\infty q^2(u)\rmd u$ and $q(x) \tilde g(\tilde r,x) = -\tilde r \int_x^\infty q(u) \tilde f(\tilde r,u) \, \rmd u \;$. Similarly, for the gap one finds:
\begin{eqnarray}
\tilde{p}_{\rm typ}(\tilde{r}|x)=\frac{2^{1/3}}{\pi}{\cal F}_{2,s}&\Bigg[&R_s\left(\left(-\tilde{r}+\frac{R_s}{q^2_s}\right)\tilde{f}_s^2-2\frac{q'_s}{q_s}\tilde{f_s}\tilde{g_s}+\left(1-\frac{q^2_s}{\tilde{r}}\right)\tilde{g}_s^2\right)\nonumber \\
&&-\frac{q^2_s}{\tilde{r}^2}\tilde{g}_s^2\Bigg] \, , \label{result_p_typ_contraint}
\end{eqnarray}
where we haved now used the shortand notations for the shifted quantities (hence the subscript $s$) $q_s=q(x-\tilde{r})$, $R_s=R(x-\tilde{r})$, ${\cal F}_{2,s}={\cal F}_2(x-\tilde{r})$, $\tilde{f}_s=\tilde{f}(-\tilde{r},x-\tilde{r})$, $\tilde{g}_s=\tilde{g}(-\tilde{r},x-\tilde{r})$.
In Fig. (\ref{rhoedge_pgap_contraint}) we show a plot of the constrained quantities: $\tilde{\rho}_{\rm edge}(\tilde{r}|0)$ in the left panel and $\tilde{p}_{\rm typ}(\tilde{r}|0)$ in the right panel when $\lmax=\sqrt{2N}$ ie $x=0$ for three different values of $\beta$ ($\beta=1$ in red, $\beta=2$ in green and $\beta=4$ in blue) computed by sampling $2. 10^7$ independent tridiagonal random matrices of size $100\times100$ \cite{DE02} for each $\beta$ and only kept the events when $|\lmax-\sqrt{2N}|<0.1$. Both panels show a good agreement with our predictions in (\ref{result_rho_edge_contraint}, \ref{result_p_typ_contraint}) for $\beta=2$ (green solid line) \cite{PS14}.

\section{Conclusion}

To conclude, we have studied the phenomenon of ``crowding'' of the eigenvalues near the largest eigenvalue $\lmax$ 
for random matrices belonging to the Gaussian $\beta$-ensembles (\ref{jPDF}). In particular we focused on the DOS (\ref{eq:def_rho_eigen}), which
is the density of eigenvalues seen from the largest eigenvalue, and the PDF of the first gap, these two quantity being related for
any finite $N$ through the relation (\ref{id_gap_DOS}). Based on exact results obtained for Hermitian
matrices, corresponding to $\beta = 2$, obtained in \cite{PS14}, and using heuristic arguments we obtained a general description
of these two quantities in the large $N$ limit. We also presented results of numerical simulations supporting our arguments. As we have seen in section \ref{sec:appli}, these quantities $\rho_{\rm DOS}(r,N)$ and $p_{\rm GAP}(r,N)$ enter naturally into the computation of physical observables in the fully connected spherical spin-glass model and it will be particularly interesting to explore further the implications of our results to the relaxational dynamics of this model. Another application of the techniques developed in \cite{PS14}, which provided a detailed analysis of an orthogonal polynomial system initially introduced in \cite{NM11}, concerns the level curvature distribution at the soft edge of random Hermitian matrices which also involves the same orthogonal polynomials \cite{Fyo12}. Finally, it will be interesting to extend the present study to other ensembles of random matrices, like the Laguerre-Wishart ensemble. Hence we hope that these results will motivate further studies of near extreme eigenvalues. 

\vspace*{0.5cm}

{\bf Acknowledgments.}
We acknowledge support by ANR grant 2011-BS04-013-01 WALKMAT and in part by the Indo-French 
Centre for the Promotion of Advanced Research under Project~$4604-3$. GS acknowledges support from Labex-PALM (Project Randmat). We also thank Y. V. Fyodorov for useful discussions.


\end{document}